%Paper: hep-th/9311097
%From: imbimbo@vxcrna.cern.ch
%Date: Wed, 17 Nov 1993 16:51:47 +0100

\input harvmac

\def\np{Nucl. Phys.}
\def\pl{Phys. Lett.}
\def\prl{ Phys. Rev. Lett.}
\def\cmp{ Comm. Math. Phys.}

\def\btd{\bigtriangledown}
\def\gmunu{g_{\mu \nu}}
\def\sgmunu{\psi_{\mu \nu}}
\def\gammu{\gamma^{\mu}}
\def\gamnu{\gamma^{\nu}}
\def\sgtildemn{{\tilde \psi}_{\mu \nu}}
\def\sginvtildern{{\tilde \psi}^{\nu \rho}}
\def\sgmixtildemr{{\tilde \psi}^{\mu}_{\rho}}
\def\ginvnr{g^{\nu \rho}}
\def\cmu{c^{\mu}}
\def\cnu{c^{\nu}}
\def\diff{\delta_c}
\def\diffg{\delta_{\gamma}}

\def\brs{s}

\def\obzero{\Omega^{(0)}}
\def\obone{\Omega^{(1)}}
\def\obtwo{\Omega^{(2)}}
\def\omone{\omega^{(1)}}
\def\omzero{\omega^{(0)}}
\def\nobzero{\Omega^{(0)}_n}
\def\nobone{\Omega^{(1)}_n}
\def\nobtwo{\Omega^{(2)}_n}
\def\twoobtwo{\Omega^{(2)}_2}
\def\epsmu{\epsilon_{\mu \nu}}
\def\det{\sqrt g}

\def\bobar{{\bar b}_0}
\def\bominus{ b_0^{-}}

\def\qbrs{Q_{BRS}}
\def\zbar{{\bar z}}
\def\Hc{{\cal H}}
\def\Pg1{{\cal P}_{g,1}}
\def\Pgn{{\cal P}_{g,n}}
\def\Modg{{\cal M}_g}
\def\Mg1{{\cal M}_{g,1}}
\def\Mgn{{\cal M}_{g,n}}
\def\S1{|\Sigma_g,P,z_P>}
\def\mutilde{{\tilde \mu}_{\psi_1,...,\psi_n}}
\def\pullmu{\mu_{\psi_1,...,\psi_n}}
\def\L0minus{L_0^-}

\overfullrule=0pt
\parskip=0pt plus 1pt

{\nopagenumbers
\font\bigrm=cmb10 scaled\magstep1
\rightline{CERN-TH-7084/93}

\rightline{GEF-Th-21/1993}
\medskip
\centerline{\bigrm ON THE SEMI-RELATIVE CONDITION FOR}

\centerline{\bigrm  CLOSED (TOPOLOGICAL) STRINGS}
\vskip 1truecm
\centerline{C. M. Becchi, R. Collina}
\vskip 5pt
\centerline{\it Dipartimento di Fisica, Universit\`a di Genova}
\centerline{\it I.N.F.N., Sezione di Genova}
\centerline{\it Via Dodecaneso 33, I-16146 Genova, Italy}
\vskip 8pt
\centerline{C. Imbimbo\footnote*{On leave from I.N.F.N., Sezione
di Genova, Italy.}}
\vskip 5pt
\centerline{\it C.E.R.N., Geneva 23, CH-1211, Switzerland}
\bigskip
\centerline{ABSTRACT}

We provide a simple lagrangian interpretation of the meaning of the
$b_0^-$ semi-relative condition in closed string theory.
Namely, we show how the semi-relative condition is equivalent to the
requirement that physical operators be cohomology classes of the BRS
operators acting on the space of local fields {\it covariant} under
world-sheet reparametrizations. States trivial in the absolute BRS
cohomology but not in the semi-relative one are explicitly
seen to correspond to BRS variations of operators which are not globally
defined world-sheet tensors. We derive the covariant
expressions for the observables of topological gravity. We use
them to prove a formula that equates the expectation value of the
gravitational descendant of ghost
number 4 to the integral over the moduli space of the Weil-Peterson
K\"ahler form.

\ \vfill

\leftline{CERN-TH 7084/93}

\leftline{GEF-Th 21/1993}

\leftline{November 1993}
\eject     }

\beginsection 1. Introduction

Physical states in closed string theory are identified with
classes of the {\it semi-relative} BRS cohomology.
The semi-relative BRS cohomology is given by equivalent classes of
elements  in the state space $\Hc$ of the string quantized on the infinite
cylinder in the conformal gauge.
A given class in this cohomology is identified by a state $|\psi>$
satisfying  the $\bominus$ condition
\eqn\semirelative{(b_0 - \bobar )|\psi> = 0,}
\noindent and annihilated by the
BRS operator $\qbrs$
modulo $\qbrs |\lambda>$, for any state $|\lambda>$ annihilated
by $\bominus \equiv b_0 -\bobar $. In the above formulae,
$b_0$ and $\bobar$ are the operators that correspond to the zero
modes of the antighost fields $b(z)$ and ${\bar b}({\bar z})$.
 The usual BRS cohomology,
consisting of $\qbrs$-invariant
states modulo $\qbrs |\lambda>$ for arbitrary $|\lambda>$, is
known as
the {\it absolute} string cohomology. Therefore the
{\it relative} cohomology is the
cohomology of the BRS operator acting on the subspace of the states
annihilated by both $b_0$ and $\bobar$.

The validity of the condition \semirelative\ has been
argued in \ref\nelson{P. Nelson, \prl\ {\bf 62} (1989) 993.}\ and
\ref\distler{J. Distler and P. Nelson, \cmp\ {\bf 138} (1991) 273.},
in the context of the so-called ``operator formalism'' \ref\gaume{
L. Alvarez-Gaum\`e, C. Gomez, G. Moore and C. Vafa, \np\
{\bf B303} (1988) 455; C. Vafa, \pl\ {\bf B190} (1987) 47.}.
Let us briefly outline this argument.

The central idea of the operator formalism is that, given a conformal
field theory with $c=0$, (e.g. a string background), one can
associate to each Riemann surface $\Sigma_g$ of genus $g$,
with a marked point $P\in \Sigma_g$ and a local coordinate system
$z_P$ centered around $P$, a state $\S1$ in
the canonical state space $\Hc$. $\S1$ is (formally) identified
with the vacuum wave functional of the string action on
$\Sigma_g/D$, where $D$ is the disk centered in $P$ with boundary at
$z_P =1$.
The triple $X\equiv (\Sigma_g,P,z_P)$ defines a point in $\Pg1$,
the space of Riemann surfaces with a single marked point and a given
local complex coordinate system around $P$. One similarly
defines the spaces $\Pgn$. In this case, the vacuum functional
on $\Sigma_g$ with $n$ disks deleted defines a state in the tensor
product ${\Hc}^{\otimes n}$. It is convenient to think of $\Pg1$
as an infinite-dimensional fiber bundle over  $\Mg1$, the (finite-dimensional)
moduli space of Riemann surfaces of genus $g$ with one puncture.
Thus, the string functional integral giving the vacuum functional
defines a (formal) map from
$\Pg1$ to $\Hc$ (and analogously defines maps from $\Pgn$ to
${\Hc}^{\otimes n}$).

Given this map, one can define the ``correlation functions'' on $\Sigma_g$ of
the ``operators''  $\psi_1,...,\psi_n$ corresponding to the states
$|\psi_1>,...,|\psi_n>$ in $\Hc$, inserted at the points $P_1,...,P_n$ of
$\Sigma_g$, as the numbers
\eqn\correlation{<\psi_{1(P_1,z_1)} ...\psi_{1(P_n,z_n)}> \equiv
<\psi_1|\otimes ...\otimes <\psi_n|\Sigma_g ,P_1,z_1,...,P_n,z_n>.}
In string theory, however, the objects of interest are not the
{\it functions} over
$\Mgn$ defined in Eq.\correlation , but top {\it forms} on $\Mgn$ which
can be integrated over the moduli space to give string amplitudes.
One can easily construct a (3g-3+n)-form
$\mutilde$ on $\Pgn$ in terms of the map
from $\Pgn$ to ${\Hc}^{\otimes n}$ discussed above, by using
the Beilinson-Konsevitch action of the Virasoro algebra
on the augmented moduli space $\Pgn$
\ref\konsevi{M.L. Konsevitch,
Funct. Anal. Appl. {\bf 21} (1988) 156.},\ref\beil{A.A. Beilinson,
Y. Manin and Y.A. Schechtman, Springer Lecture
Notes in Mathematics {\bf 1289} (1987) 52.}:

\eqn\Pgform{\eqalign{\mutilde (V_1,\ldots, V_{3g-3+n},
{\bar V}_1,\ldots,{\bar V}_{3g-3+n})& =
\left( <\psi_1|\otimes \ldots \otimes <\psi_n|\right)
b(v_1){\bar b}({\bar v}_1)\ldots \cr
\ldots & b(v_{3g-3+n}){\bar b}({\bar v}_{3g-3+n})
|\Sigma_g ,P_1,z_1,\ldots,P_n,z_n>,\cr}}

\noindent where $V_i$ (${\bar V}_i$), with $i=1,\ldots,3g-3+n$, are tangent
(anti)holomorphic vectors to $\Pgn$ at
the point $(\Sigma_g,P_1,z_1,\ldots,P_n,z_n)$; $v_i$ (${\bar v}_i$)
are corresponding vector fields on $\Sigma$, (anti)holomorphic on the unit
disk $D$ minus the point $P$ and meromorphic at $P$;
$b(v_i) \equiv
\oint_P dz_P b_{zz} v_i^z$ are the associated anti-ghost insertions.

The problem with $\mutilde$
is that it is a form on $\Pgn$, rather than on $\Mgn$.
By selecting a section $\sigma$ of $\Pgn$ over $\Mgn$ -- i.e.
by choosing local coordinate systems $(z_{P_1},\ldots,z_{P_n})$ that
vary smoothly as
$(\Sigma_g,P_1,\ldots,P_n)$ varies over $\Mgn$ -- one obtains a $3g-3+n$
form $\pullmu$ on $\Mgn$ by pulling back $\mutilde$ via $\sigma$:
\eqn\pullback{\pullmu = \sigma^*\mutilde.}
$\pullmu$ depends on $\sigma$, that is on the chosen family of
local coordinate systems $z_{P_i}$. However, standard arguments
show that if the states $|\psi_i>$ are BRS closed then $\mutilde$
is a closed form on $\Pgn$, and the cohomology class of the pulled
back $\pullmu$ is independent on $\sigma$ \nelson .

As pointed out in \nelson , the problem with Eq.\pullback\ is
that a {\it global} section $\sigma$
does not exist (for generic genus $g$).
The best that one can do is to choose $\sigma$ to be a section continuous
up to a phase. For the corresponding $\pullmu$ to be continuous
on $\Mgn$ it is necessary that $\mutilde$  be invariant under
the transformation corresponding to
$L_0 -{\bar L}_0\equiv \L0minus$ and that $\mutilde$
annihilates tangent vectors in the direction of this Virasoro generator. This
is the case precisely when the physical states $|\psi_i>$ satisfy the
$\bominus$ condition of Eq.\semirelative .

To summarize, the derivation of the semi-relative condition \semirelative\
in the operator formalism is striclty Hamiltonian. ``Correlators'' of
operators on higher genus Riemann surfaces are defined via scalar products
in the canonical state space $\Hc$. Such definition involves the formal map
from the infinite-dimensional bundle $\Pg1$ to the canonical space
of states. It relies on some geometrical properties of this bundle --
namely, on the Beilinson-Konsevitch Virasoro action upon it and on the
absence of a global section.
The fact that states trivial in
the absolute BRS cohomology but non-trivial in the semi-relative cohomology
do not in general decouple is, ultimately, a self-consistency requirement of
the
operator formalism: in this approach a consistent definition of string
amplitudes would not even be possible, were the
$\bominus$ condition not fulfilled.

However, it should be clear that in a lagrangian field
theoretical approach to first quantized closed strings,  the
relevant BRS cohomology
is neither a matter of choice nor of definition.
Given a reparametrization invariant lagrangian
on a two-dimensional arbitrary closed surface, the physical
states of the corresponding closed string are automatically
determined by the gauge invariance of the theory.

Here, in fact, we show that in the {\it covariant} lagrangian approach
to first quantized closed bosonic strings and
topological strings, the semi-relative
condition \semirelative\ is identical to the requirement that physical
observables be covariant under coordinate reparametriza\-tions of the
world-sheet. In other words, we prove that the physical states
in the semi-relative cohomology are in one-to-one
correspondence with non-trivial
classes of the BRS operator acting on the space of {\it covariant}
local field operators.
The fact that operators corresponding to states
trivial in the absolute cohomology but not trivial in the
semi-relative one do not decouple in physical amplitudes
is the direct consequence of the non-vanishing of world-sheet integrals of
two-forms which are locally but not globally exact.
Therefore, the ``equivariance'' principle of closed string
theory is simply that of
covariance under general coordinate transformations.
This in turn is dictated by the fact that string amplitudes
are integral over the world-sheet of correlators of field operators.

Beyond the virtue of
simplicity, our novel derivation of the well known condition
\semirelative\ also has the advantage of leading to covariant expressions
for the operatorial observables in question. We expect that
the covariant expressions we derive for the observables of topological
strings will be important for the explicit field theoretical
calculation of their correlators, still lacking in the literature.
We intend to report on this in a future work.

\beginsection 2. Closed bosonic critical string

Let $\gmunu$ be the two-dimensional metric on the world-sheet,
$\cmu$ the ghost fields, $X$ the matter fields. We will denote
by $\brs$ the BRS operator acting on the fields. Since in critical
strings the Liouville field is not dynamical,
gauge transformations are two-dimensional diffeomorphisms
accompanied with a compensating Weyl transformation. Therefore
the action of $\brs$ is :
\eqn\critbrs{\eqalign{
\brs \gmunu &= \diff \gmunu - (D c)\gmunu \cr
\brs \cmu &= {1\over 2} \diff \cmu = \cnu \partial_{\nu} \cmu \cr
\brs X &= \diff X,\cr}}

\noindent where $\diff$ denotes the action of the diffeomorphisms
with parameters $\cmu$, and $Dc \equiv D_{\alpha} c^{\alpha}$,
with $D_{\alpha}$ being the covariant derivative.
Let $d$ be the exterior derivative on
forms; $\brs$ and $d$ commute among themselves.
To each two-form $\obtwo$, $s$-closed modulo $d$,  one can associate a
one-form
$\obone$ and a zero-form $\obzero $ satisfying  the famous descent equations:
\eqn\descent{\eqalign{
\brs \obtwo &= d \obone \cr
\brs \obone &= d \obzero \cr
\brs \obzero &= 0.\cr }}
Furthermore, if $\obzero$ is a local observable, $\obtwo$ belongs to the
cohomology of $s$ modulo $d$.
Noting that the cohomology of $s$ in the space of the two and one-forms
is trivial, we can revert the statement above: if $\obtwo$ belongs to the
cohomology of $s$ modulo $d$, $\obzero$ is necessarily in the cohomology of
$s$.

Now take
\eqn\two{\obtwo = {1\over 2} \det R \epsmu dx^{\mu}\wedge dx^{\nu}
\equiv {\cal R}^{(2)}}
where $g=det (\gmunu)$, $R$ is the scalar two-dimensional curvature and
$\epsmu$ is the antisymmetric numeric tensor defined by
$\epsilon_{12}=1$.
One can verify that $\obtwo$ satisfies the descent equations \descent\ with
$\obzero$ and $\obone$ given by:
\eqn\solution{\eqalign{
\obone &=\det \epsmu (\cnu R + \ginvnr \partial_{\rho}(Dc))dx^{\mu}
\cr
\obzero &= \det \epsmu ({1\over 2}\cmu \cnu R + \cmu \ginvnr
\partial_{\rho} (Dc)).\cr }}
$\obzero$ is the local operator which corresponds to the dilaton state at zero
momentum $|D> = (c_{-1}c_1 - {\bar c}_{-1} {\bar c}_1 )|0>,$ as it is
apparent by considering $\obzero$ in the conformal gauge $\gmunu =
e^{\phi}\delta_{\mu \nu}$ and applying it to the $SL(2,C)$ invariant
vacuum $|0>$. $|D>$ is the unique example in critical string of a state
which is trivial in the absolute $\qbrs$ cohomology (since
$|D> = \qbrs c_0^- |0>$) but non-trivial in the semi-relative one
(since $\bominus (c_0^- |0>) \ne 0$), and which therefore does not
decouple in generic amplitudes.

It is now clear how this translates into our covariant
field theoretical framework.
If $\obtwo= d \omone$, then it follows easily from the descent equations that
\eqn\trivone{\obone = \brs \omone + d \omzero}
for some $\omzero$, and
\eqn\trivtwo{\obzero = \brs \omzero.}
However if $\omone$ cannot be chosen to be
a covariant one-form, $\omzero$ will not be a
scalar under reparametrizations, as it is implied  by Eq. \trivone .
This is precisely the case for the
operators corresponding to the dilaton state. $\obtwo$ in Eq.\two\ is
locally but not globally exact, that is ${\cal R}^{(2)} = d\omone$
but $\omone$ is not a globally defined one-form.
Therefore one can write $\obzero$ in
Eq. \solution\ as the $\brs$ variation of something,
$\obzero = \brs \omzero$,
but $\omzero$ will not be coordinate independent. Moreover, the triviality
of the $s$-cohomology in the space of the two and one-forms guarantees the
absence of any globally defined $\omzero$ satisfying Eq. \trivtwo .

The facts that
the dilaton state does not decouple in
generic physical amplitudes and that its expectation value is
topological follow directly from the covariant two-form
representation in Eq.\two . In contrast,
the calculation of the dilaton correlation functions in the conformal
operator formalism requires a careful consideration
of the transformation properties of non-covariant correlators
under changes of coordinate patches
\ref\pol{J. Polchinsky, \np\ {\bf B307} (1988) 61.},\nelson .

One can calculate explicitly $\omzero$ and $\omone$ by choosing a
particular
coordinate system. In order to compare with the conformal field theory
formalism, let us choose a system of holomorphic coordinates with
metric given by $ds^2 = |\lambda( dz + \mu d\zbar)|^2,$ where $\mu$
is a Beltrami differential. Then

\eqn\one{ \omone = {2\over \Theta}\left(\partial\mu
- \mu\bar{\partial}\bar{\mu}
+ {1\over 2}\left({\bar\btd} - \mu\btd\right)ln\Theta\right)d\zbar
- c.c.,}

\noindent where we introduce the symbol $\Theta \equiv 1 - \mu\bar\mu$
and the derivative
$\btd \equiv \partial - \bar{\mu}\bar{\partial}$, and where $c.c.$
denotes the conjugate expression in which all quantities are substituted
with their barred expressions.
For $\omzero$ one obtains the result:

\eqn\noscalar{\omzero = \btd c + {\bar c}(\omone)_{\zbar}  - c.c..}

Going to a conformal frame with $\mu=0$, $\omzero$ reduces to
$\partial c - {\bar \partial}{\bar c}$, the conformal operator
creating the state $c_0^-|0>$ which does
not satisfy the semi-relative condition \semirelative .

\beginsection 3. Topological gravity

The basic fields of topological gravity \ref\per{
J. Labastida, M. Pernici and E. Witten, \np\ {\bf B310} (1988) 611.}
are the metric $\gmunu$ and
its gravitino superpartner $\sgmunu$, together with the anticommunting
ghost $\cmu$ and their commuting superpartners $\gammu$. Note that
in this case the Liouville field is a dynamical field and thus
is not inert under BRS transformations. The nilpotent BRS
transformations are:
\eqn\topbrs{\eqalign{
\brs \gmunu &= \diff \gmunu + \sgmunu \cr
\brs \sgmunu &= \diff \sgmunu - \diffg\gmunu \cr
\brs \cmu &= {1\over 2} \diff \cmu  + \gammu = \cnu \partial_{\nu} \cmu  +
\gammu \cr
\brs \gammu &= {1\over 2} \diff \gammu- {1\over 2}\diffg \cmu=
\cnu \partial_{\nu} \gammu - \partial_{\nu}\cmu \gamnu , \cr}}
where $\diff$ and $\diffg$ are the variations under reparametrizations
with parameters $\cmu$ and $\gammu$ respectively.

Let us solve again the descent equations \descent\ with
$\obtwo$ given by the Euler two-form as in Eq. \two\ and the BRS
transformations defined by Eq.\topbrs . For the one and
zero-forms one obtains the following expressions:
\eqn\topsolution{\eqalign{
\obone &=\det \epsmu (\cnu R + D_{\rho} (\sginvtildern - {1\over 2}g^{\rho
\nu}\psi^{\sigma}_{\sigma})\,)dx^{\mu} \cr
\obzero &= \det \epsmu ({1\over 2}\cmu \cnu R + \cmu D_{\rho}
(\sginvtildern - {1\over 2}g^{\rho \nu}\psi^{\sigma}_{\sigma}) +
D^{\mu}\gamnu  - {1\over 4} \sgmixtildemr \sginvtildern),\cr }}
where $\sgtildemn\equiv \sgmunu - {1\over 2}\gmunu \psi^{\sigma}_{\sigma}$
is the traceless part of the gravitino. These formulas for
$\obone$ and $\obzero$ are the covariant generalizations of the
expressions which hold in the conformal gauge and which are given
in \ref\verlinde{
E. Verlinde and H. Verlinde, \np\ {\bf B348} (1991) 457.}.

As before, since $\obtwo = d\omone$ is locally exact,
it follows from the descent equation that
$\obzero = \brs \omzero$,  but with $\omzero$ not being
a reparametrization invariant scalar. Using a holomorphic coordinate
system parametrized by the Beltrami differential $\mu$, one
can compute $\omzero$  to be:
\eqn\local{\omzero =  \btd c + {\bar c}(\omone)_{\zbar} +
{{\bar \mu}\psi \over \Theta } - c.c.,}
where $\psi = \brs \mu$ and ${\bar \psi}=\brs {\bar \mu}$
are the components of the traceless part
of the gravitino field $\sgtildemn$ in the holomorphic coordinate system.
In a conformal frame in which $\mu =0,$ one can
again check that $\omzero$ reduces
to $\partial c - {\bar \partial}{\bar c}$, the conformal
operator creating the state $c_0^- |0>$ which violates the semi-relative
condition in Eq.\semirelative .

Thanks to
the commutative nature of the superghosts $\gammu$,
of topological gravity, one can construct an infinite
set of non-vanishing
observables $\nobzero= (\obzero )^n$ with $n= 0,1,\ldots$ by taking positive
powers of the ``dilaton'' operator $\obzero$ \verlinde .  The operators
$\nobzero$ are believed to exhaust the BRS local cohomology.
They all create states which
are trivial in the absolute BRS state cohomology but non-trivial in the
semi-relative one.  The one-form and two-form expressions associated to
$\nobzero$ can be expressed in terms of the operators $\obzero$, $\obone$ and
$\obtwo$:
\eqn\nforms{\eqalign{\nobone = & n \, {\obzero}^{n-1} \obone \cr \nobtwo= &
n \, {\obzero}^{n-1}\obtwo + {1\over 2} n (n-1) {\obzero}^{n-2}\obone
\wedge \obone  .\cr}}

It is expected that the word-sheet integral of the matrix elements of the
two-form $\nobtwo$ be a closed $2n-2$-form on the moduli space $\Modg$. In
particular, $<\int_{\Sigma_g}\obtwo_2>$ should correspond to a closed
$(1,1)$ form on $\Modg$.

We intend to present a more detailed analysis of the correlation functions
of the integrated observables $\nobtwo$ in a covariant lagrangian
approach in a forthcoming paper. Here we limit ourselves to
the following observation.

When considering vacuum expectation values of the integrated observables
$\nobtwo$, all the terms in Eq.\topsolution\ but the term bilinear in
the gravitino field
average to zero. The gravitino field $\tilde{\psi}^{\mu \nu}$ is naturally
identified with a cotangent vector on $\Modg$ \per \nref\witten{E. Witten,
\np\ {\bf B340} (1990) 281.}-\ref\singer{L. Baulieu and I. Singer,
\cmp\ {\bf 135} (1991) 253.}, and one can show that after performing
the fermionic functional integration,
it can be substituted with following expression:
\eqn\cotangent{\tilde{\psi}^{\mu \nu}= {1\over
\sqrt{g}}\partial_i(\sqrt{g}g^{\mu \nu})dm^i \equiv {1\over \sqrt{g}}d_m
(\sqrt{g}g^{\mu\nu}),}
where $m_i$ are coordinates on $\Modg$, $\partial_i$ and $dm^i$ are the
corresponding derivatives and one-form elements. $g^{\mu \nu} =
g^{\mu \nu}(x;m^i)$ is a two-dimensional metric which represents the gauge
equivalence class of metrics corresponding to the point of $\Modg$
with coordinates $\{ m^i\}$.
On account of the explicit formulas given in Eq. \topsolution\ above,
the vacuum expectation value of $\twoobtwo$ is reduced therefore
to the following closed
$(1,1)$ form on the moduli space $\Modg$:
\eqn\closed{\mu_{\Modg} \equiv <\int_{\Sigma} \twoobtwo > =
\int_{\Sigma}d^2x g \epsilon_{\alpha \beta}\left(-{1\over 2}R
{\tilde \psi}^{\alpha \mu}{\tilde \psi}^{\beta}_{\mu} + D_{\mu} {\tilde
\psi}^{\mu \alpha} D_{\nu}{\tilde \psi}^{\nu \beta}\right).}

One can verify that the two-form in Eq.\closed , when evaluated using a
constant curvature section $g^{\mu\nu}(x;m^i)$, coincides with the
Weil-Peterson K\"ahler two-form on $\Modg$
\ref\weil{L.V. Ahlfors, J, Analyse Math. {\bf 9} (1961) 161;
S. Wolpert, Ann. of Math. {\bf 117} (1983) 207.}.
Hence the expectation value of $3g-3$ integrated $\Omega_2^{(2)}$'s
should formally be given by the volume of $\Modg$ computed via the
Weil-Peterson symplectic form. To be able to integrate over $\Modg$
powers of the $\mu_{\Modg}$ given in Eq.\closed ,
one needs to study the problem of extending $\mu_{\Modg}$ to
${\overline \Modg}$, the Deligne-Mumford compactification of $\Modg$.
Because of the divergencies of the Weil-Peterson
form at the boundary of $\Modg$ \ref\masur
{H. Masur, Duke Math. J., {\bf 43} (1976) 623;
S. Wolpert, Amer. J. Math., {\bf 107} (1985) 1485.}, this is
a delicate issue  and corresponds in the field theoretical language
to the problem of possible BRS anomalies
induced by contributions at the boundary of $\Modg$
\ref\cec{M. Bershadsky, S. Cecotti, H. Ooguri and C. Vafa,
Harvard preprint, HUTP-93/A008, hep-th/9302103.}.

\listrefs

\bye